\documentclass[aps,twocolumn,showpacs,preprintnumbers,amsmath,amssymb,superscriptaddress,nofootinbib,english]{revtex4-1}
\usepackage{times,amsmath,amsfonts,amssymb,epstopdf}
\usepackage{graphicx}
\usepackage{dcolumn}
\usepackage{bm}
\usepackage{epsfig}
\usepackage{hyperref}
\usepackage[usenames]{color}
\usepackage{url}
\usepackage[normalem]{ulem}
\usepackage[T1]{fontenc}
\usepackage[dvipsnames]{xcolor}

\def\be{\begin{equation}}
\def\ee{\end{equation}}
\def\ba{\begin{eqnarray}}
\def\ea{\end{eqnarray}}
\begin{document}

\title{Spatial variations of the fine-structure constant in symmetron models}

\author{Marvin~F.~Silva}
\email[Email address: ]{marvinf.silva@gmail.com}
\affiliation{Centro de Astrofisica, Universidade do Porto, Rua das Estrelas, 4150-762 Porto, Portugal}

\author{Hans~A.~Winther}
\email[Email address: ]{h.a.winther@astro.uio.no}
\affiliation{Institute of Theoretical Astrophysics, University of Oslo, 0315 Oslo, Norway}

\author{David~F.~Mota}
\email[Email address: ]{d.f.mota@astro.uio.no}
\affiliation{Institute of Theoretical Astrophysics, University of Oslo, 0315 Oslo, Norway}

\author{C.~J.~A.~P.~Martins}
\email[Email address: ]{Carlos.Martins@astro.up.pt}
\affiliation{Centro de Astrofisica, Universidade do Porto, Rua das Estrelas, 4150-762 Porto, Portugal}

\date{\today}

\begin{abstract}
We investigate the variation of the fine-structure constant, $\alpha$, in symmetron models using N-body simulations in which the full spatial distribution of $\alpha$ at different redshifts has been calculated. In particular, we obtain simulated sky maps for this variation, and determine its power spectrum. We find that in high-density regions of space (such as deep inside dark matter halos) the value of $\alpha$ approaches the value measured on Earth. In the low-density outskirts of halos the scalar field value can approach the symmetry breaking value and leads to significantly different values of $\alpha$. If the scalar-photon coupling strength $\beta_\gamma$ is of order unity we find that the variation of $\alpha$ inside dark matter halos can be of the same magnitude as the recent claims by Webb {\em et al.} of a dipole variation. Importantly, our results also show that with low-redshift symmetry breaking these models exhibit some dependence of $\alpha$ on lookback time (as opposed to a pure spatial dipole) which could in principle be detected by sufficiently accurate spectroscopic measurements, such as those of ALMA and the ELT-HIRES.\end{abstract}


\maketitle


\section{Introduction}

These are exciting times for cosmology and particle physics. They both have successful standard models, which are in
agreement with a plethora of experimental and observational data. Nevertheless, there are also strong hints that neither of these models is complete. In particular, the observational evidence for the acceleration of the universe \cite{SN1,SN2} (which presently adds up to several tens of standard deviations, if all available data is combined) implies the existence of new, currently undiscovered physics. The question is then what new degrees of freedom may be relevant, and what consistency tests can be used to confirm their presence.

After a quest of several decades, the recent LHC detection of a Higgs-like particle \cite{ATLAS,CMS} finally confirms that fundamental scalar fields are part of Nature's building blocks. A pressing follow-up question is whether the associated field has a cosmological role, or indeed if there is another cosmological counterpart. Regardless of the answer to these questions scalar fields are ubiquitous as explanations for a range of theoretical paradigms in cosmology---including possible explanations for the acceleration itself.

Moreover, when a new dynamical degree of freedom such as a scalar field is responsible for the recent acceleration, one can show \cite{Carroll} that if it couples to the rest of the model (which it will naturally do, unless one postulates a new symmetry principle to suppress these couplings) it will also lead to variations of nature's dimensionless fundamental couplings, which one can hope to detect through direct astrophysical or local (laboratory) measurements \cite{Book}. There have been several claims of variations of the fine-structure constant $\alpha$ at the parts per million level, culminating in the recent evidence for a dipole in the variation \cite{Dipole1,Dipole2}. If confirmed, then this is direct evidence of new physics. 

Do the fundamental constants vary? In addition to its intrinsic relevance, answering this question has key implications for cosmology and fundamental physics, and in particular can shed light on the enigma of dark energy \cite{Amendola,euclid,koivisto,review}. An ESO-VLT Large Program, whose data analysis is ongoing \cite{LP1,LP2}, is trying to clarify this issue, but an unambiguous answer may have to wait until a new generation of high-resolution ultra-stable spectrographs such as PEPSI, ESPRESSO and ELT-HIRES is available. Moreover, a resolution demands not only better data, but also independent ways to search for these variations, which may confirm or contradict these indications.

In this paper we will investigate whether the tentative claims of $\alpha$ variations can be explained in the context of scalar-tensor modifications of gravity by looking at a particular modified gravity model, the symmetron.

The simplest model that produces a variation of $\alpha$ is obtained by promoting the fine-structure constant to a scalar field via the  field-strength tensor $F_{\mu\nu}^2 \to f(\phi)F_{\mu\nu}^2$. A spacetime variation of $\phi$ will then induce a variation of $\alpha$ \cite{mb1,mb2,mb3}. 

In the symmetron model \cite{symmoriginal,olivepospelov}, the vacuum expectation value (VEV) of a scalar field depends on the local mass density, becoming large in regions of low density, and small in regions of high density. The coupling of the scalar to matter is proportional to the VEV and this leads to a viable theory where the scalar can couple with gravitational strength in regions of low density, but is decoupled and screened in regions of high density. This is achieved through the interplay of a symmetry breaking potential and a universal quadratic coupling to matter. In vacuum, the scalar acquires a VEV which spontaneously breaks the $\mathcal{Z}_2$ symmetry $\phi\to-\phi$. In the regions of sufficiently high matter density, the field is confined near $\phi=0$, and the symmetry is restored. The fifth force arising from the matter coupling is proportional to $\phi$ making the effects of the scalar small in high density regions.

The cosmology of coupled scalar field models is usually strongly constrained by local gravity experiments, which put limits on the range and the coupling strength of the scalar field \cite{symmcosm2,ms,li}. For the symmetron this restricts the Compton wavelength of the scalar field to be less than a few megaparsec in vacuum. There do exist cases in which signatures on the linear perturbations are found \cite{symmcosm1,symmnbody}, but in most cases the signatures are found in the non-linear regime.

In \cite{symmnbody,systsim} the effects on non-linear structure formation using N-body simulations was investigated. Such studies have shown that the fifth-force leads to an enhancement of the matter power-spectrum on non-linear scales and in the low mass tail of the halo mass-function. Another interesting signature found in the model, and other scalar-tensor modified gravity models, is an environmental dependence of observables \cite{envdep1}. In \cite{envdep} a significant difference between the lensing and dynamical masses of dark matter halos was found in the symmetron model which depends on both the halos mass and environment. 

The key feature in such scalar-tensor theories which leads to the environmental dependence of certain physical observables is the clustering and the spatial inhomogeneities of the scalar degree of freedom. The later, due to the coupling to baryons and dark matter, becomes inhomogeneous at scales of its Compton wavelength. Within the framework of varying alpha modes, this was computed in \cite{mb1} in the liner regime and in \cite{mb2,mb3}  in the nonlinear regime of structure formation. Spatial inhomogeneities in the Gravitation constant, $G$, were calculated in \cite{clifton}.

The setup of this paper is as follows. In Sec.~\ref{review} we give a brief review of the symmetron model, in Sec.~\ref{analysis} we present the results from the analysis of the N-body simulations and in Sec.~\ref{alphapk} we present the numerically determined power spectrum for the $\alpha$ variations and compare it with an analytic estimate. Finally in Sec.~\ref{conc} we present the conclusions. In this paper we use units of $c=1$ throughout.


\section{The Symmetron Model}\label{review}
In this section we give a brief review of the symmetron model. This is not meant to be exhaustive, but only to describe the aspects that will be relevant for our analysis. We refer the reader to the literature already cited above for a more detailed description.
\\\\
The symmetron model is a scalar-tensor modification of gravity described by the action
\begin{align}
S &= \int dx^4\sqrt{-g}\left[\frac{R}{2}M_{\rm pl}^2 - \frac{1}{2}\left(\partial\phi\right)^2 - V(\phi)\right] \nonumber\\
&+ S_m(\Psi_m; g_{\mu\nu}A^2(\phi))
\end{align}
where $g = \det g_{\mu\nu}$, $M_{\rm pl} = 1/\sqrt{8\pi G}$, $S_m$ is the matter-action and we have used units of $\hbar = c \equiv 1$. The matter fields $\Psi_m$ are coupled to the scalar field via a conformal coupling
\begin{align}
\tilde{g}_{\mu\nu} = g_{\mu\nu}A^2(\phi)
\end{align}
Because of this coupling the matter-fields will experience a fifth-force, which in the non-relativistic limit is given by
\begin{align}
\vec{F}_{\phi} \equiv \frac{dA(\phi)}{d\phi}\vec{\nabla}\phi = \frac{\phi\vec{\nabla}\phi}{M^2}
\end{align}
where the last equality only holds for the symmetron. For the symmetron the potential is chosen to be of the symmetry breaking form
\begin{align}
V(\phi) = - \frac{1}{2}\mu^2\phi^2 + \frac{1}{4}\lambda\phi^4
\end{align}
where $\mu$ is a mass-scale and the conformal coupling is chosen as the simplest coupling consistent with the potential symmetry $\phi\to-\phi$
\begin{align}
A(\phi) =  1 + \frac{1}{2}\left(\frac{\phi}{M}\right)^2
\end{align}
where $M$ is a mass-scale and $\lambda$ a dimensionless coupling constant. A variation of the action with respect to $\phi$ gives the field-equation
\begin{align}
\nabla^2\phi = \frac{dV_{\rm eff}}{d\phi}
\end{align}
The dynamics of $\phi$ is determined by the effective potential
\begin{align}
V_{\rm eff} &= V(\phi) + A(\phi)\rho_m\nonumber\\
&= \frac{1}{2}\left(\frac{\rho_m}{\mu^2M^2} - 1\right)\mu^2\phi^2 + \frac{1}{4}\lambda\phi^4
\end{align}
In the early Universe where the matter-density is high the effective potential has a minimum at $\phi = 0$ where the field will reside. As the Universe expands the matter density dilutes until it reaches a critical density $\rho_{\rm SSB} = \mu^2M^2$ for which the symmetry breaks and the field moves to one of the two new minima $\phi = \pm \phi_0 = {\mu}/\sqrt{\lambda}$.
\\\\
The fifth-force between two test-particles residing in a region of space where $\phi = \phi_{\rm local}$ can be found to be
\begin{align}
\frac{F_{\phi}}{F_{\rm gravity}} =  2\beta^2 \left(\frac{\phi_{\rm local}}{\phi_0}\right)^2,~~~~~~\beta = \frac{\phi_0 M_{\rm pl}}{M^2}
\end{align}
for separations within the Compton wavelength $\lambda_{\rm local} = 1/\sqrt{V_{\rm eff,\phi\phi}(\phi_{\rm local})}$ of the scalar-field. For larger separations the force is suppressed by a factor $e^{-r/\lambda_{\rm local}}$. In the cosmological background before symmetry breaking $\phi_{\rm local} \approx 0$ and the force is suppressed. After symmetry breaking the field moves towards $\phi = \pm\phi_0$ and the force can be comparable with gravity for $\beta = \mathcal{O}(1)$. In high density regions, like the Sun and our Galaxy, non-linear effects in the field-equation ensure that the force is effectively screened thereby evading local gravity constraints.
\\\\
In the following discussion it will be convenient to introduce the variables
\begin{align}
a_{\rm SSB} &= \left(\frac{\rho_{m0}}{\rho_{\rm SSB}}\right)^{1/3}\\
\lambda_{\phi 0} &= \frac{1}{\sqrt{2}\mu}
\end{align}
together with the already defined quantities
\begin{align}
\beta &= \frac{\phi_0 M_{\rm pl}}{M^2}\\
\rho_{\rm SSB} &= \mu^2M^2
\end{align}
Here $\beta$ is the coupling strength relative to gravity, $\rho_{\rm SSB}$ is the density in at which the symmetry is broken, $a_{\rm SSB}$ is the corresponding scale-factor for when this happens in the cosmological background and $\lambda_{\phi 0}$ is the range of the fifth-force when the symmetry is broken. Local gravity constraints \cite{symmoriginal,symmnbody,symmcosm1,symmcosm2} force the range of the field to satisfy
\begin{align}
\lambda_{\phi 0} \lesssim \text{Mpc}/h
\end{align}
for symmetry breaking close to today, i.e. $a_{\rm SSB} \sim 1$.

 
\subsection{Coupling $\phi$ to electromagnetism}
The electromagnetic field is unaffected by a conformal transformation because of the conformal invariance of the EM action, $S_{\rm EM}(A_{\mu}; g_{\mu\nu}A^2(\phi)) \equiv S_{\rm EM}(A_{\mu}; g_{\mu\nu})$. We can however consider generalizations where the EM field is coupled to the scalar field via
\begin{align}
S_{\rm EM} =  -\int dx^4\sqrt{-g} A^{-1}_{\gamma}(\phi)\frac{1}{4}F_{\mu\nu}^2
\end{align}
With this coupling we still have that perfect fluid radiation does not affect the Klein-Gordon equation for the scalar field because the Stress-Energy tensor of the EM field is traceless. This coupling leads to the fine-structure constant depending on $\phi$ as
\begin{align}
\alpha = \alpha_0 A_{\gamma}(\phi)
\end{align}
where $\alpha_0$ is the laboratory value.

We will consider two different coupling functions below.
\subsubsection*{Quadratic coupling}
The simplest choice for $A_{\gamma}$, compatible with the $\phi\to-\phi$ symmetry of the symmetron, is
\begin{align}
A_{\gamma}(\phi) = 1 + \frac{1}{2}\left(\frac{\beta_{\gamma}\phi}{M}\right)^2
\end{align}
where $\beta_{\gamma}$ is the scalar-photon coupling relative to the scalar-matter coupling, i.e. a value of $\beta_\gamma = 1$ implies that the scalar-photon coupling is the same as the scalar-matter coupling. A variation of $\phi$ leads to a variation of the fine-structure constant $\alpha$ with respect to the laboratory value $\alpha_0$:
\begin{align}
\frac{\Delta \alpha}{\alpha} = A_{\gamma}(\phi)-1 = \frac{1}{2}\left(\frac{\beta_{\gamma}\phi}{M}\right)^2
\end{align}

For the symmetron we have
\begin{align}
\frac{1}{2}\left(\frac{\beta_{\gamma}\phi}{M}\right)^2 \simeq \beta^2\beta_\gamma^2\left(\frac{0.5}{a_{\rm SSB}}\right)^3&\left(\frac{\phi}{\phi_0}\right)^2\left(\frac{\lambda_{\phi 0}}{\text{Mpc}/h}\right)^2\times\nonumber\\
&\times\left(\frac{\Omega_{m0}}{0.25}\right)\times10^{-6}
\end{align}
For our fiducial model parameters $a_{\rm SSB} \sim 0.5$, $\beta \sim 1$, $\lambda_{\phi 0} \sim 1 \text{Mpc}/h$ we can have a maximum variation of alpha, achieved in the broken phase $\phi = \phi_0$, of
\begin{align}
\left.\frac{\Delta \alpha}{\alpha}\right|_{\rm max} \simeq \beta_\gamma^2 \times 10^{-6}
\end{align}
which for $\beta_\gamma \sim 1$ are close to the recent analysis by Webb et al. \cite{Dipole1,Dipole2}.
\subsubsection*{ Linear coupling}
Another possibility is the well motivated exponential coupling
\begin{align}
A_{\gamma}(\phi) = e^{\frac{\beta_\gamma\phi}{M_{\rm pl}}} \simeq  1 + \frac{\beta_\gamma\phi}{M_{\rm pl}}
\end{align}
which we have expanded as a linear function since the argument of the exponential is required by observations to be much less than unity. However, this coupling does not respect the $\phi\to-\phi$ symmetry. For the symmetron model we find
\begin{align}
\frac{\Delta \alpha}{\alpha} &= \frac{\beta_\gamma\phi}{M_{\rm pl}} \\
&= \beta_\gamma\beta\left(\frac{0.5}{a_{\rm SSB}}\right)^3\left(\frac{\phi}{\phi_0}\right)\left(\frac{\lambda_{\phi 0}}{\text{Mpc}/h}\right)^2\left(\frac{\Omega_{m0}}{0.25}\right)\times10^{-6}\nonumber
\end{align}
which for $\beta_\gamma \sim 1$ is again of the same order as found above for the quadratic coupling.

Note that in the last scenario the variation is proportional to $\phi$ instead of $\phi^2$. This means that the variation can have both signs if the symmetry is broken differently in different places in the Universe, i.e. if we have domain walls. At a naive, qualitative level, a domain wall based senario capable of accounting for the claimed dipole would simultaneouly require low tension walls (so they evade other cosmological constraints) and presumably a number of walls per Hubble volume of order unity; those two requirements are not necessarily compatible for the simplest domain wall models \cite{Walls1,Walls2}, although they may be made so with some fine-tuning \cite{Walls3,Walls4}. This will be interesting to study further, although it will require N-body simulations where the full time evolution of the scalar field is solved \cite{nbodytime}. We thus leave these issues for subsequent work, and in the rest of this paper we will only consider the quadratic coupling.


\section{Analysis of N-body simulations}\label{analysis}

We will quantify the possible variations of $\alpha$ in this class of models by resorting to N-body simulation results, in which the full spatial distribution of $\phi$ at different redshifts has been calculated. Variations of $\alpha$ have been studied in N-body simulations for the Bekenstein-Sandvik-Barrow-Magueijo model in \cite{alphanbody1,alphanbody2}. The N-body simulations for the symmetron which we use were taken from the analysis in \cite{symmnbody}.

The physical parameters used in the simulations are as follows: the present dark-energy fractional energy density $\Omega_\Lambda = 0.733$ and $\Omega_m = 0.267$, $H_0 = 71.9$ km/s/Mpc, $n_s = 0.963$ and $\sigma_8 = 0.801$. These values are consistent with the WMAP7 best-fit $\Lambda$CDM model \cite{wmap7}. The size of the simulation box is $64$ Mpc/h, in which $h = H_0/$(100km/s/Mpc) and $N=256^3$ dark matter particles was used. The background evolution in the symmetron model is very close to that of $\Lambda$CDM justifying this choice. However, the presence of a fifth-force in the simulations alters structure formation. An example of this can be seen in Fig.~(\ref{fig:mass}) where we show the mass function for our symmetron simulations (presented below) compared to $\Lambda$CDM.
\\\\
The symmetron parameters for the three simulations (all performed with the same initial density configuration) we have analyzed are\footnote{The simulations labels corresponds to those in \cite{symmnbody}.}:
\begin{align}
A: a_{\rm SSB} = 0.66, \lambda_{\phi 0} = 1.0, \beta = 1.0\\
C: a_{\rm SSB} = 0.50, \lambda_{\phi 0} = 1.0, \beta = 1.0\\
E: a_{\rm SSB} = 0.33, \lambda_{\phi 0} = 1.0, \beta = 1.0
\end{align}
Thus the three models have symmetry-breaking phase transitions at redshifts of 0.5, 1 and 2 respectively, well within the range of current optical and radio tests of the stability of fundamental couplings \cite{Book}.

The value of $\beta$ does not influence the solution to the Klein-Gordon equation directly; only indirectly in the clustering of matter. This implies that the value of $\alpha$ for a model with a different value of $\beta$ will be the same modulo differences in structure formation. In the limit $\beta \to 0$ the power-spectrum, mass-function and other clustering related observables reduce to that of $\Lambda$CDM.

\begin{figure}
\begin{center}
\includegraphics[scale=0.65]{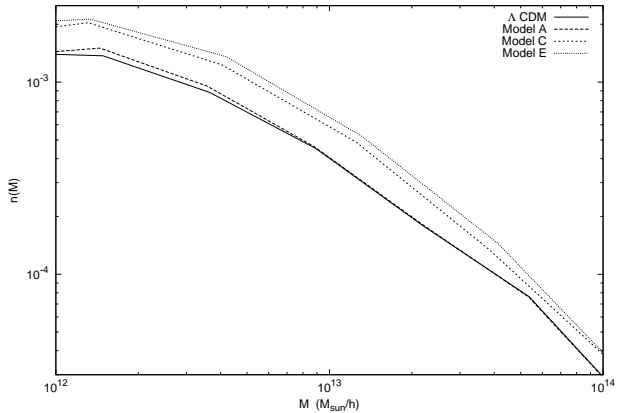}
\caption{Mass function for the symmetron models A, C and E at $z=0$. For comparison, the mass function of $\Lambda$CDM is also shown.}
\label{fig:mass}
\end{center}
\end{figure}

\subsection*{Variation of $\alpha$ inside dark matter halos}
\begin{figure}
\begin{center}
\includegraphics[scale=0.7]{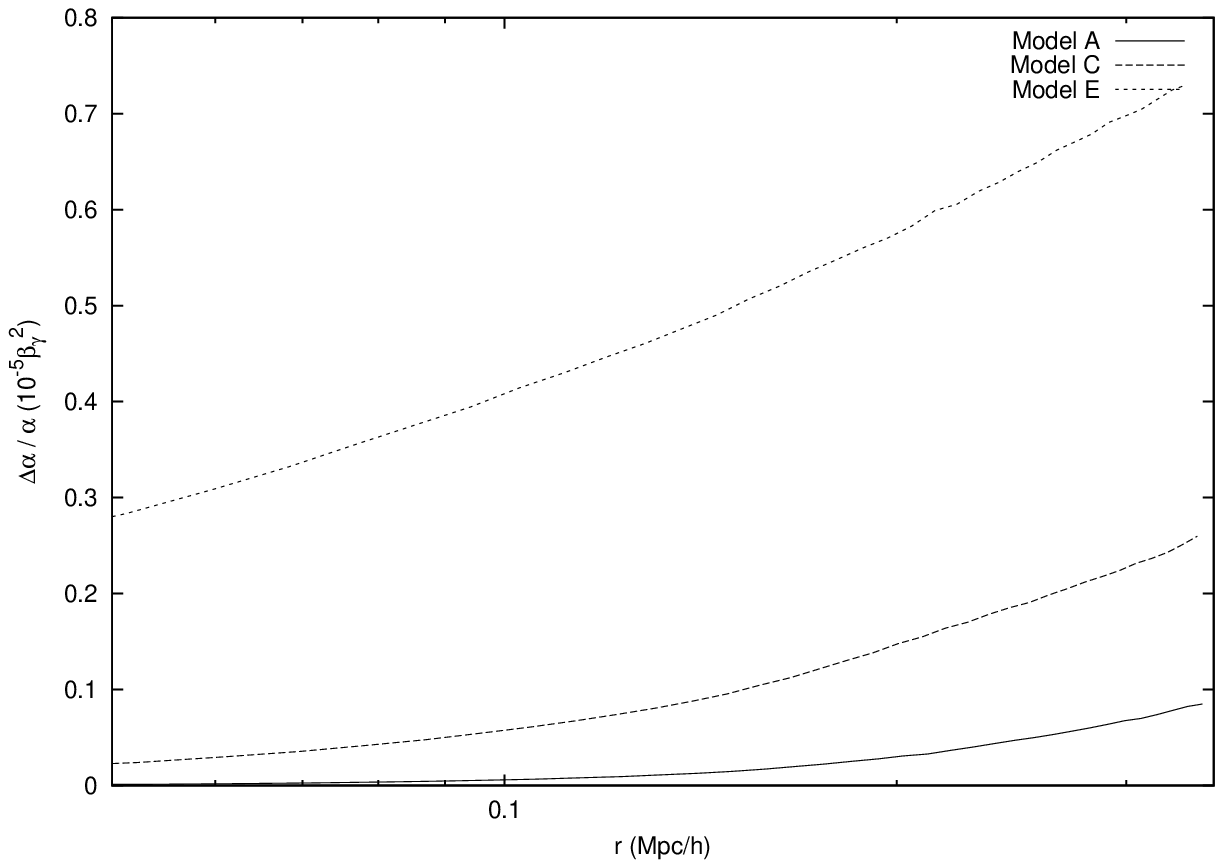}
\includegraphics[scale=0.7]{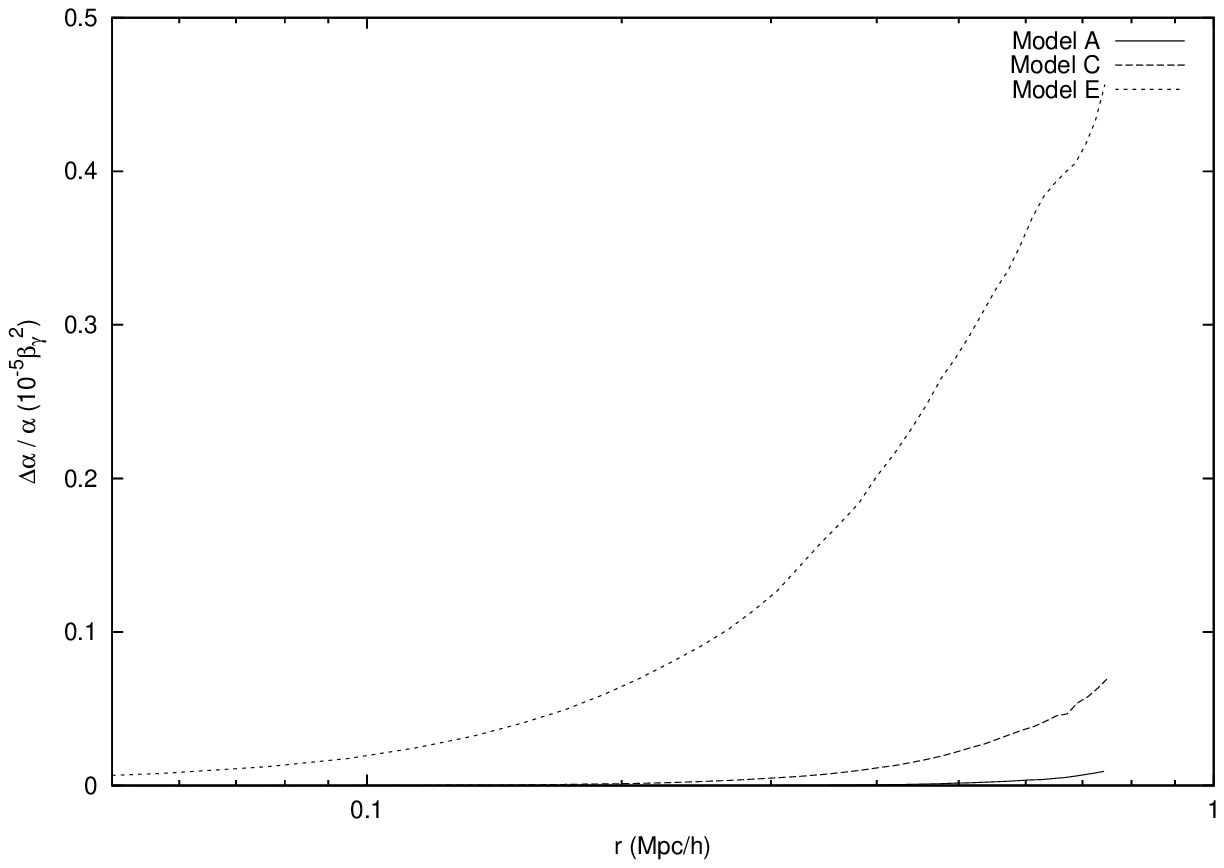}
\caption{Variation of $\alpha$ inside halos of different masses ($M = 5\cdot 10^{12}~M_{\rm sun}/h$ above and $M = 5\cdot 10^{13}~M_{\rm sun}/h$ below) for model A, C and E at $z=0$.}
\label{fig:modcomp}
\end{center}
\end{figure}

\begin{figure*}
\begin{center}
\includegraphics[scale=0.9]{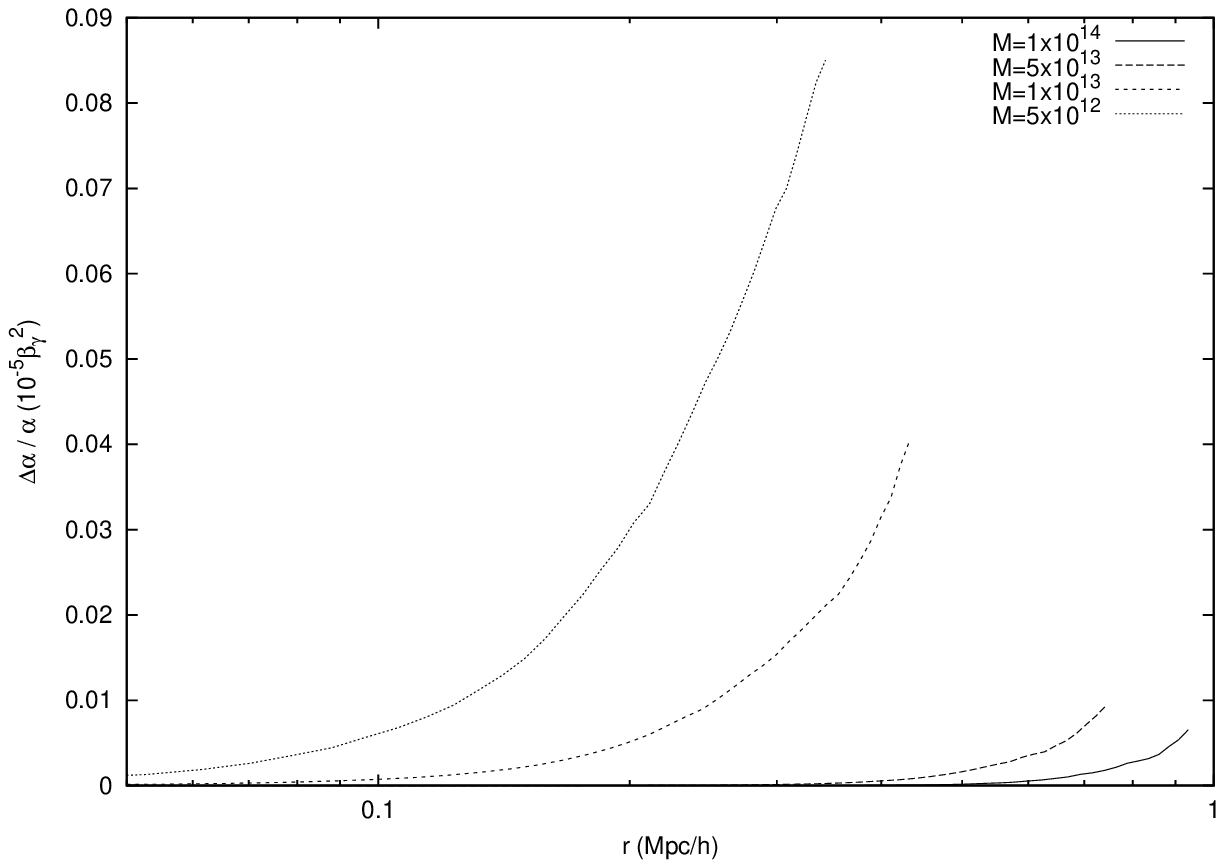}
\includegraphics[scale=0.9]{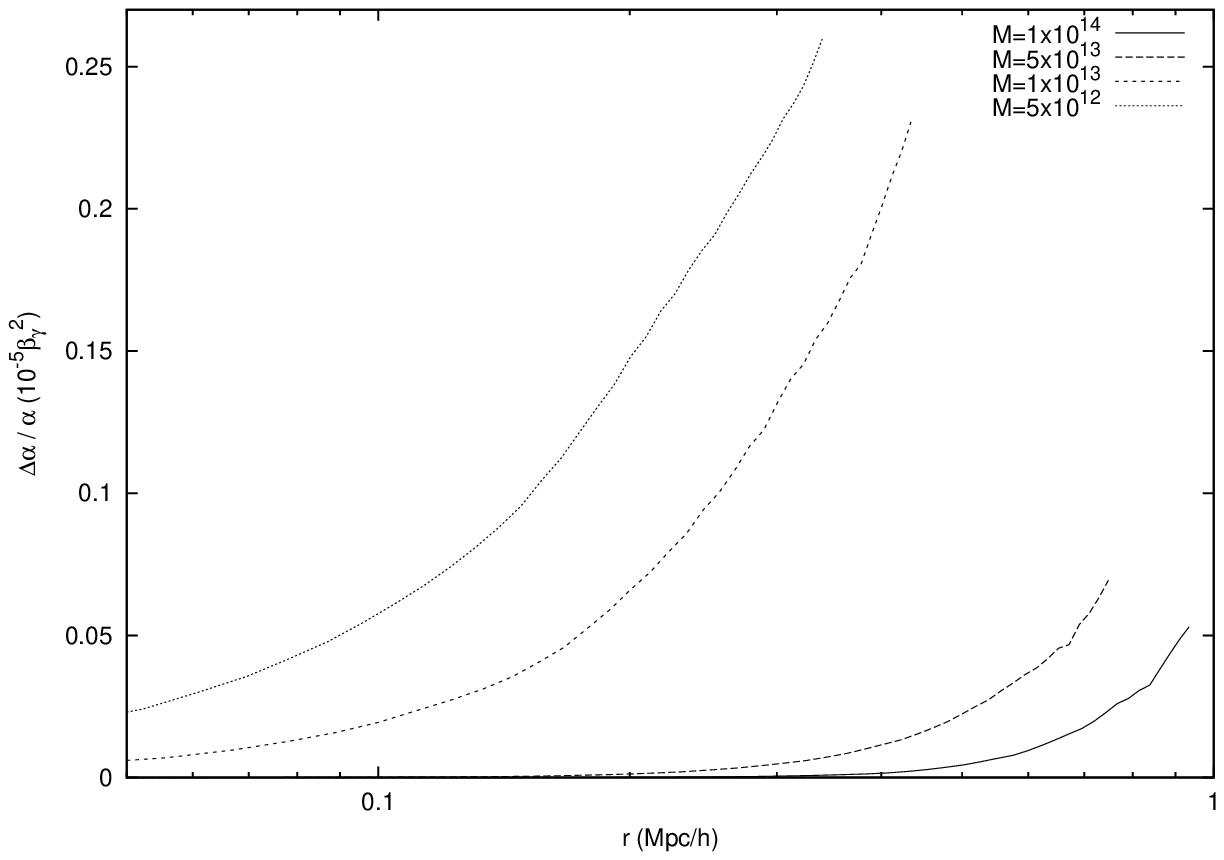}\\
\includegraphics[scale=0.9]{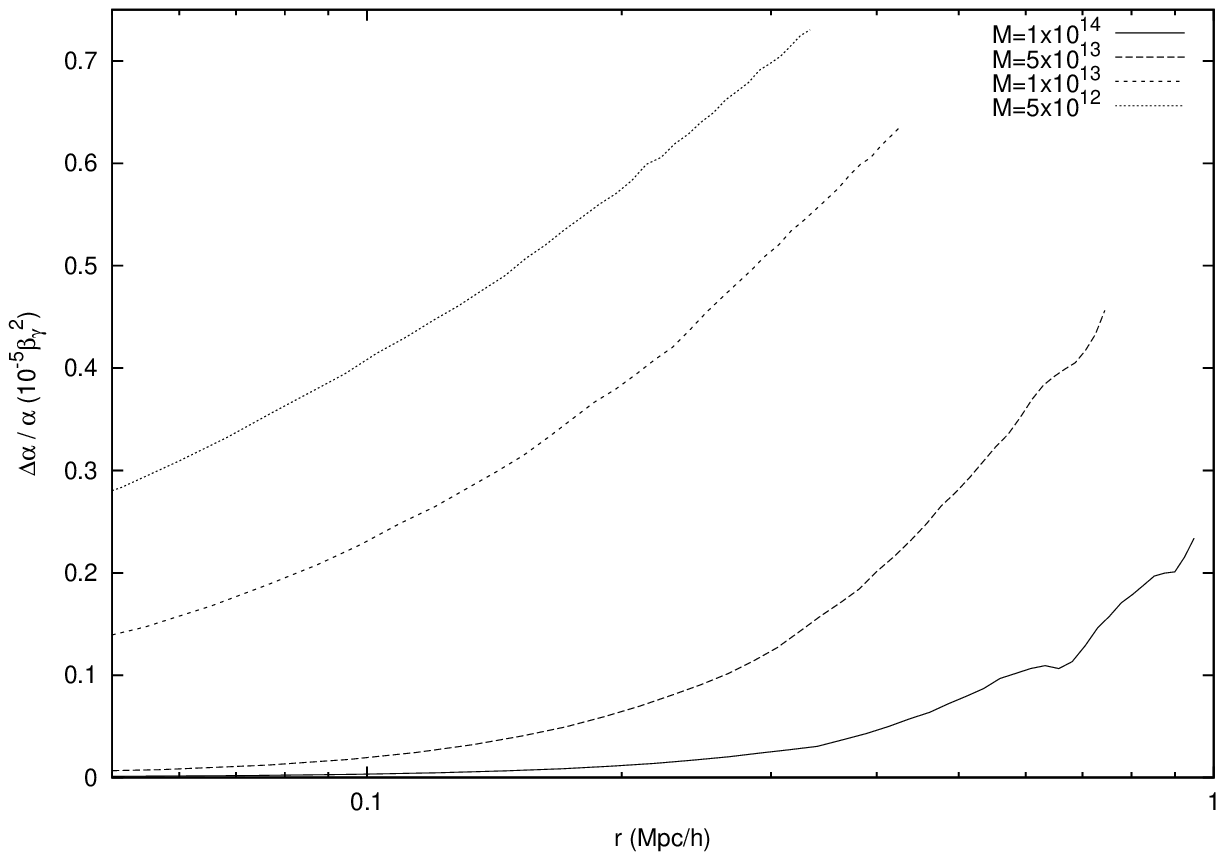}
\caption{Variation of $\alpha$ inside halos of different masses for model A (top), C (middle) and E (bottom) at $z=0$.}
\label{fig:ps0z0}
\end{center}
\end{figure*}

\begin{figure}
\begin{center}
\includegraphics[scale=0.7]{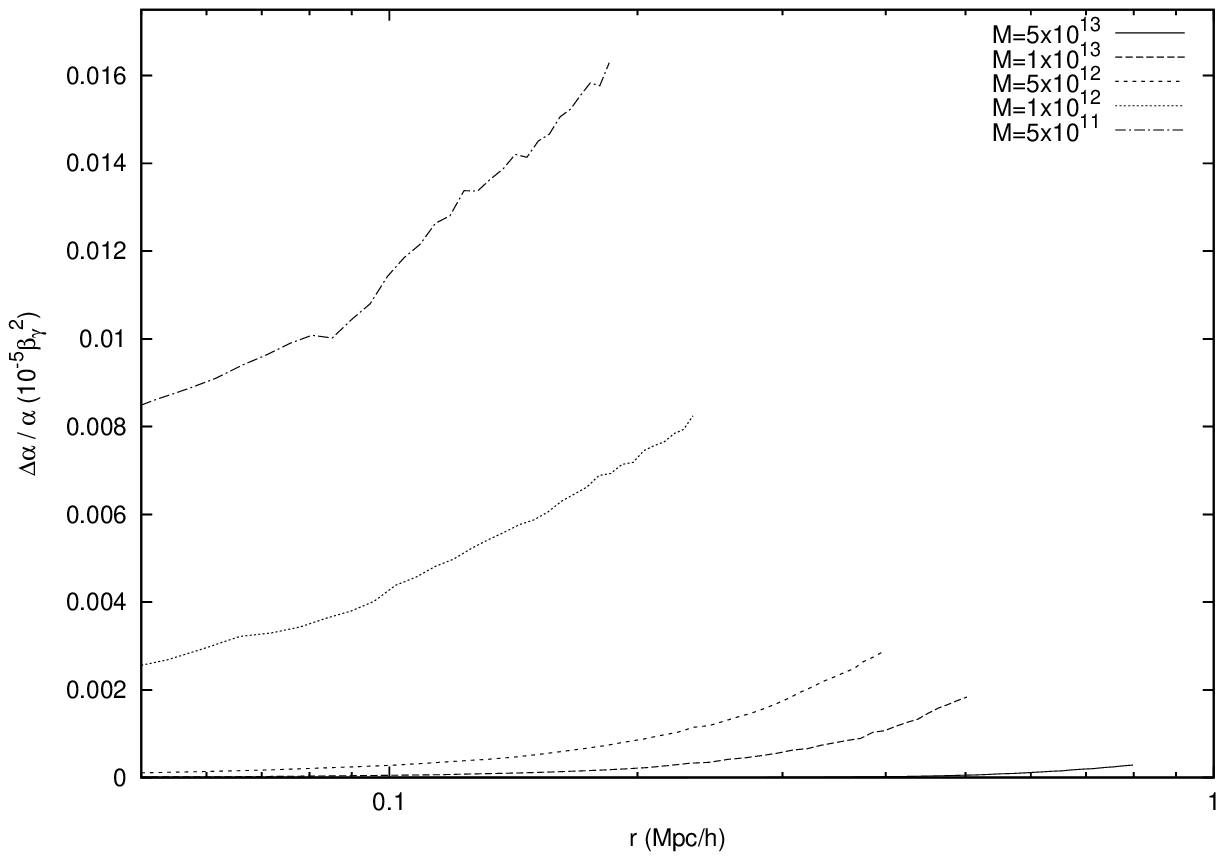}\\
\includegraphics[scale=0.7]{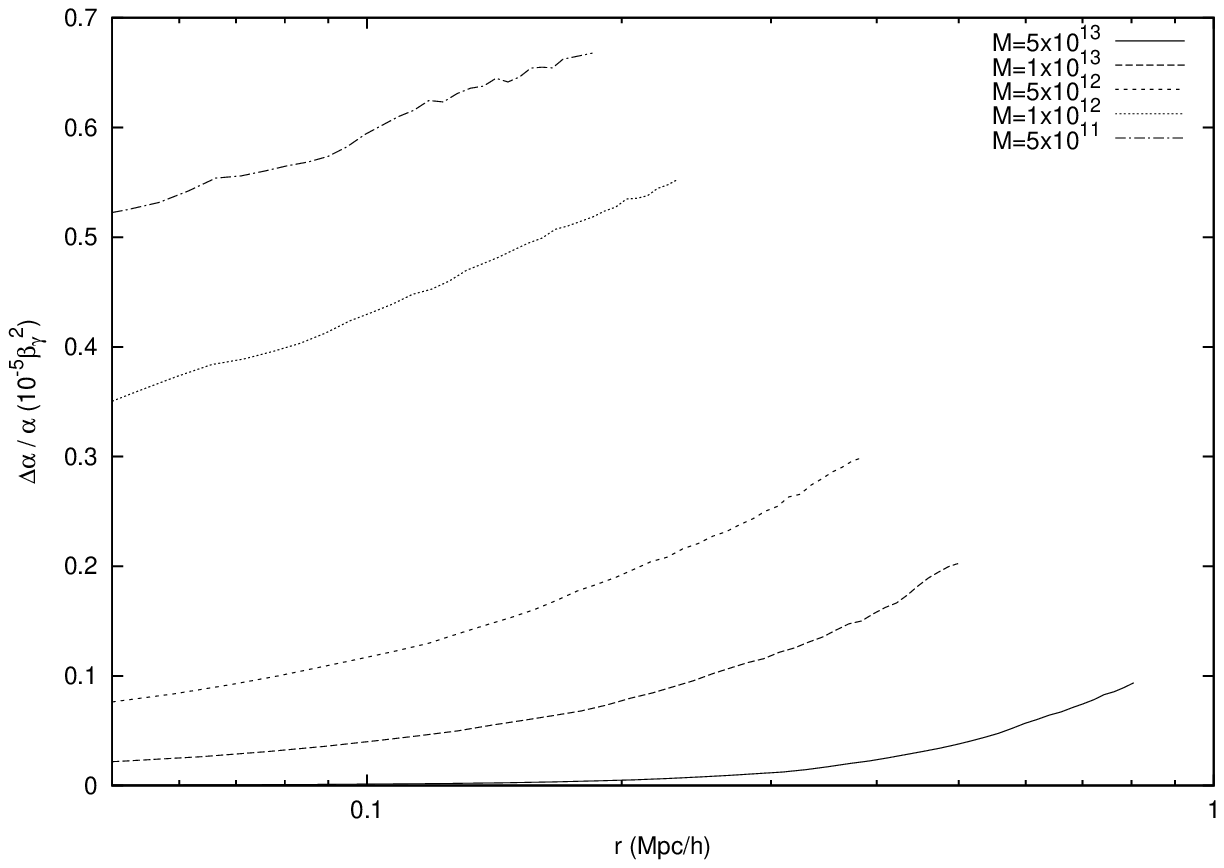}
\caption{Variation of $\alpha$ inside halos of different masses for model C (top) and E (bottom) at $z=1.0$.}
\label{fig:ps0z1}
\end{center}
\end{figure}

We first extract the position of the particles in the simulation and the corresponding scalar-field value. Then we run the halo-finder code AHF \cite{ahf} to locate the halos. Having identified the location of the halos we bin the particles belonging to the different halos and the scalar-field value to get the scalar-field profiles for halos of different sizes. 
\\\\
In Figs.~(\ref{fig:modcomp},\ref{fig:ps0z0},\ref{fig:ps0z1}) we show the halo profile inside halos of different masses at redshifts $z=0$ and $z=1$. Instead of plotting $\phi(r)$ directly we show the corresponding variation of $\alpha$ (relative to the value measured on Earth) 
\begin{align}
\frac{\Delta \alpha}{\alpha} = \frac{1}{2}\left(\frac{\beta_{\gamma}\phi(r)}{M}\right)^2
\end{align}
As expected from the screening property of the model, larger halos correspond to smaller values of $\phi$. We also see that models where the symmetry breaks early on have larger values of $\phi$. This is also as expected from the screening property; the earlier the symmetry breaks the larger is the critical density threshold for screening.
\\\\
For all the models we have considered here, the variation of $\alpha$ from inside to outside of dark matter halos is of order $\sim 10^{-5}\beta_\gamma^2$. Thus for the scalar-photon coupling $\beta_\gamma$ of order unity the variation of $\alpha$ from Earth to the outskirts of dark matter halos is of the same order of magnitude as the tentative claims by Webb \emph{et al.} \cite{Dipole1,Dipole2}. Comparison of Figs.~(\ref{fig:modcomp}) and (\ref{fig:ps0z1}) also suggests a moderate redshift dependence of the values of $\alpha$. Although currently available measurements don't have the sensitivity to seach for these effects, they should be detectable by the next generation of ultra-stable spectrographs, for example by observing lines of sight were several absportion clouds can be found \cite{LP1}.

\subsection*{Skymaps of $\Delta \alpha /\alpha$}

\begin{figure*}
\begin{center}
\includegraphics[scale=0.45]{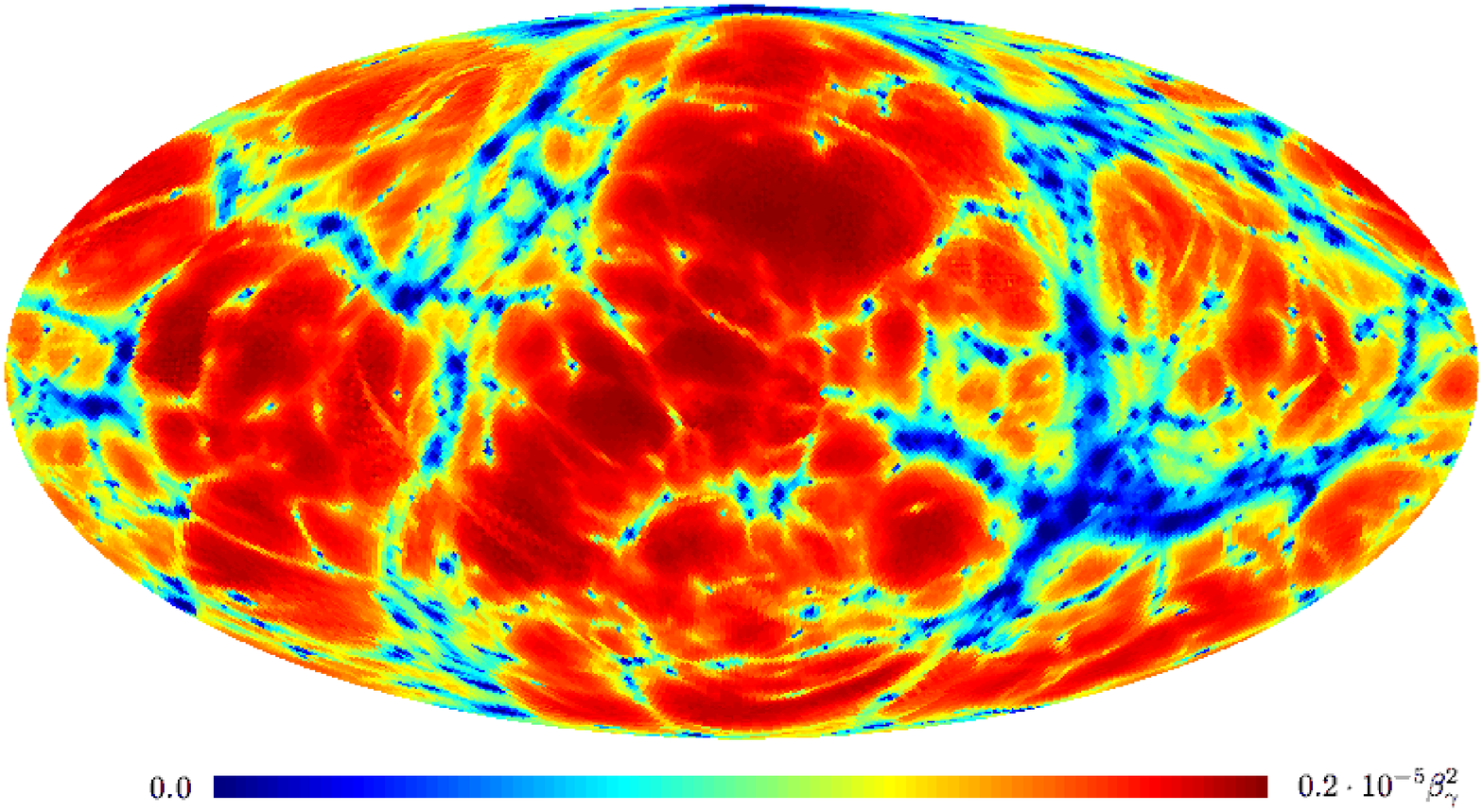}\\
\includegraphics[scale=0.45]{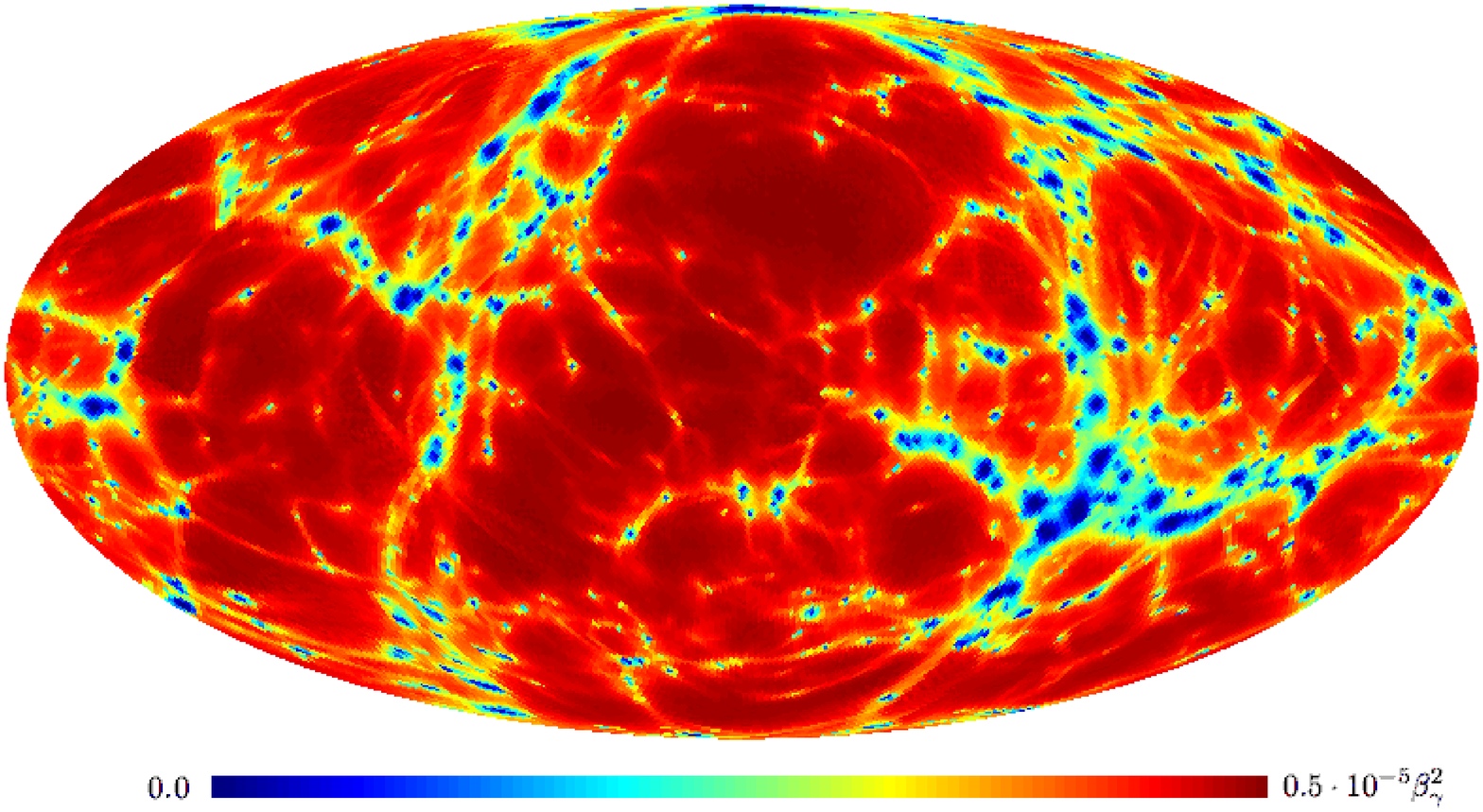}\\
\includegraphics[scale=0.45]{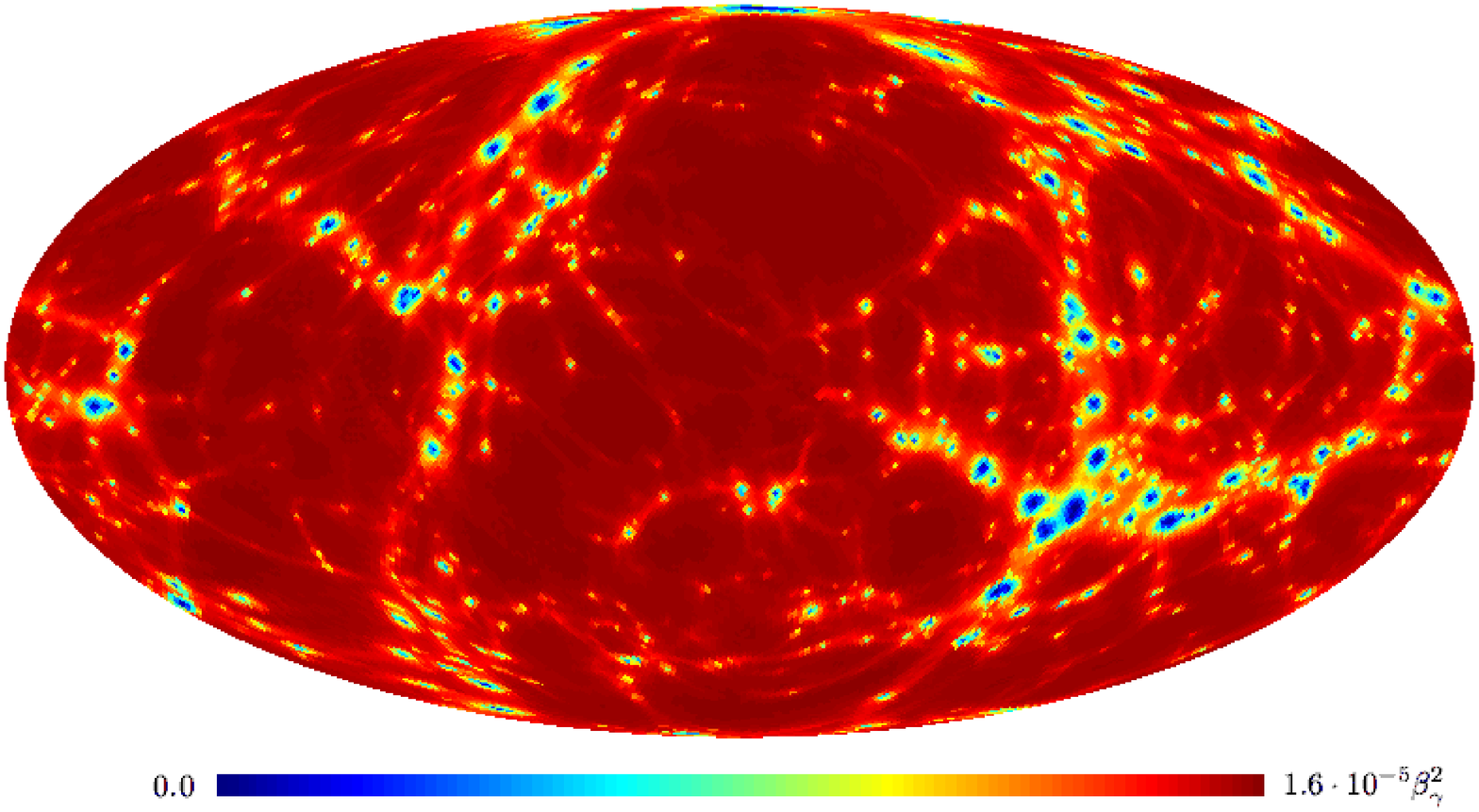}
\caption{The variation $\frac{\Delta \alpha}{\alpha}$ over the sky at $z=0$ for model A (top), C (middle) and E (bottom).}
\label{fig:map}
\end{center}
\end{figure*}

In Fig.~(\ref{fig:map}) we show the variation of $\alpha$ over the whole sky at $z=0$ for the models A, C and E. The maps are produced by first placing an observer at the center of our N-body simulation box and then projecting down the values of $\alpha$ for all particles within a sphere with co-moving radius $R=60$ Mpc$/h$ centered around the observer. In other words, the maps show the value of $\alpha$ across the whole sky for a thin redshift-slice around a given redshift.

We see a clear correlation with the time symmetry breaking takes place and the fraction of the sky where $\alpha$ deviates from its value on Earth.  For model E, symmetry breaking takes place at redshift $z_{\rm SSB} \equiv 1/a_{\rm SSB} - 1 = 2$ and almost all the sky (at $z=0$) except inside massive clusters shows a large $\Delta\alpha/\alpha$ deviation. For model A where symmetry breaking takes place at $z_{\rm SSB} = 0.5$ a larger fraction of the sky will have the same value as on Earth. In all the maps, the value of $\alpha$ is highly correlated with the matter density field as we found from the halo analysis (see also the next two sections).

\begin{figure*}
\begin{center}
\includegraphics[scale=0.45]{E.z0.eps}\\
\includegraphics[scale=0.45]{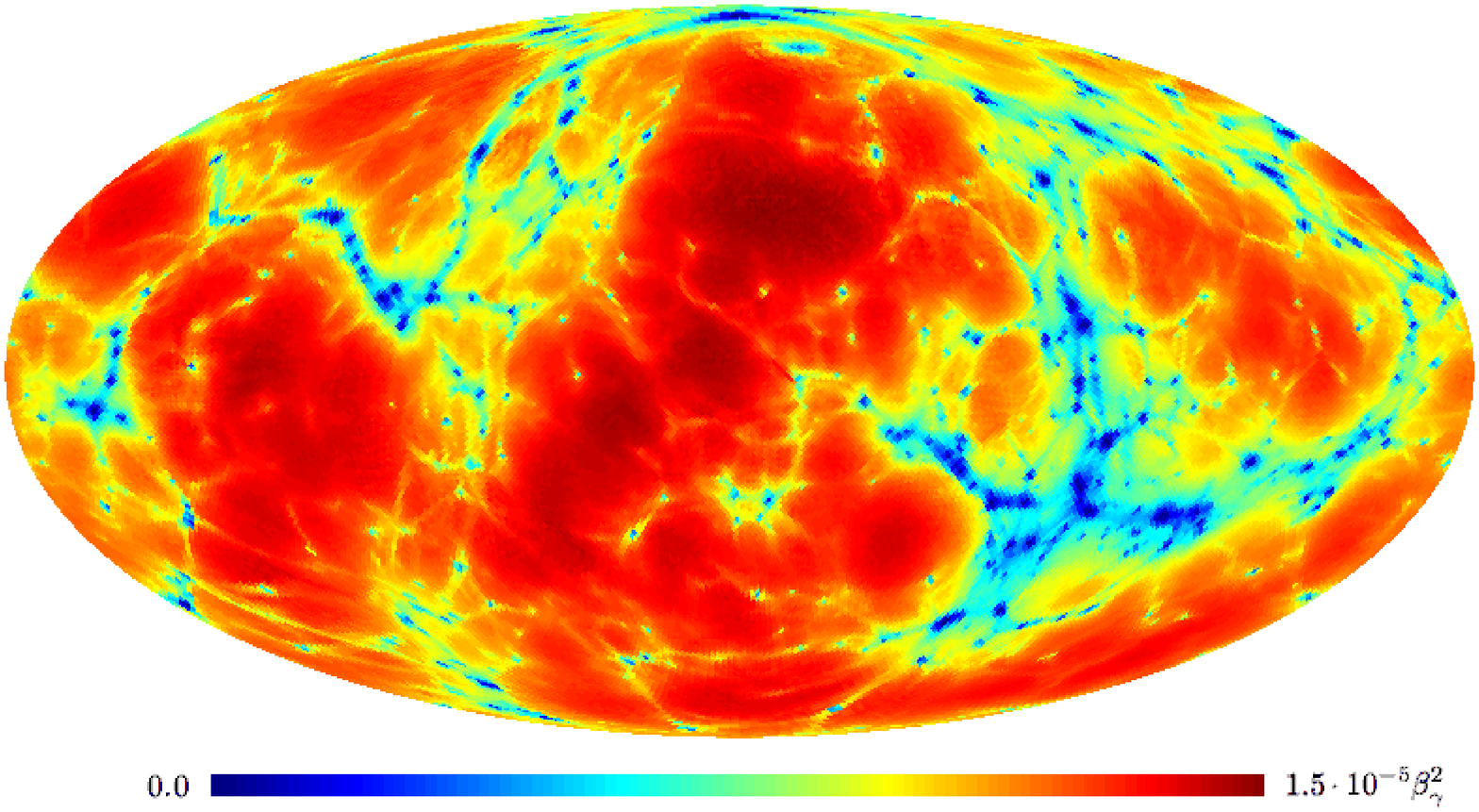}\\
\includegraphics[scale=0.45]{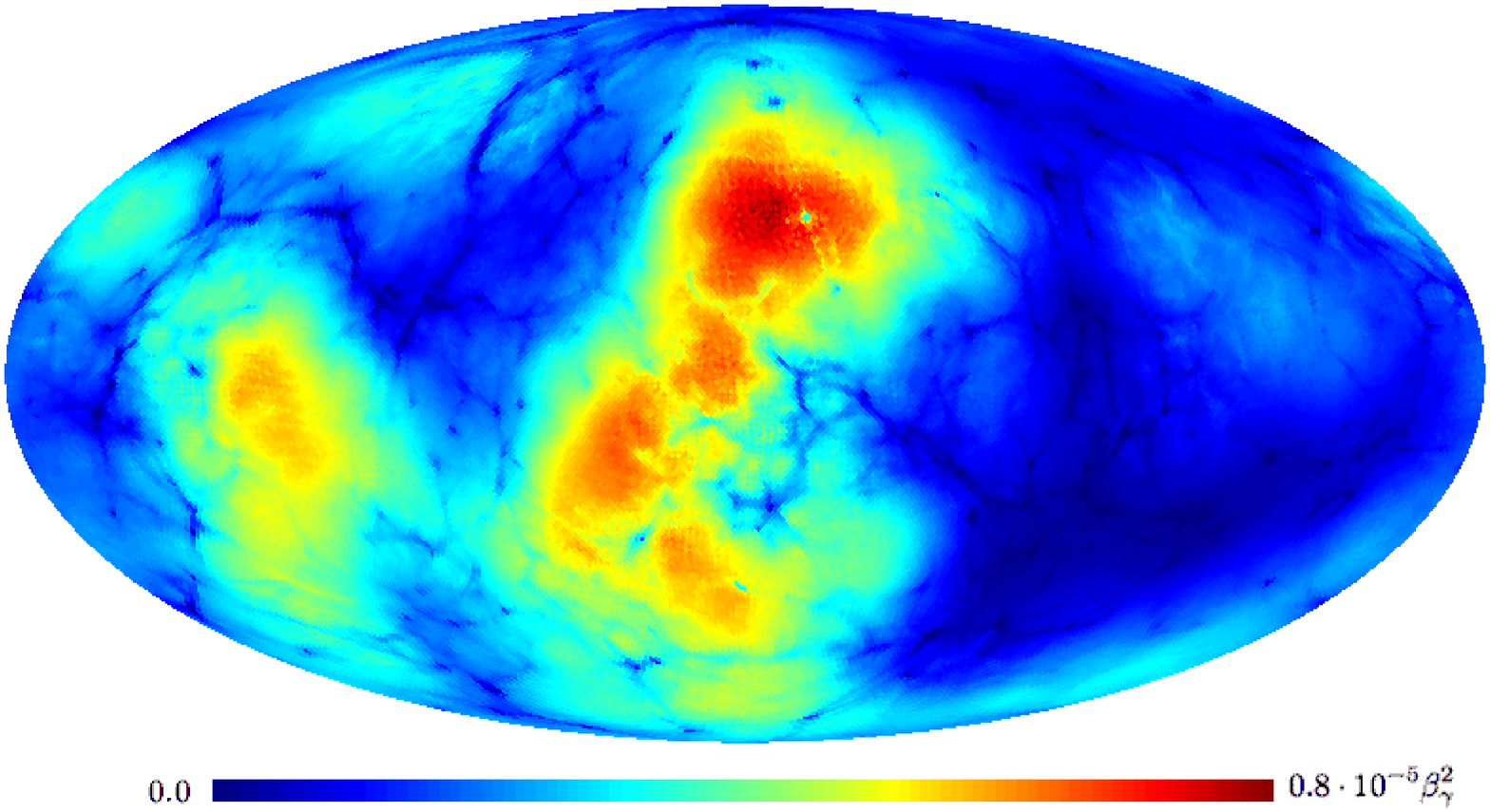}
\caption{The variation $\frac{\Delta \alpha}{\alpha}$ over the sky for model E at $z=0$ (top), $z=1$ (middle) and $z=2$ (bottom).}
\label{fig:map2}
\end{center}
\end{figure*}

In Fig.~(\ref{fig:map2}) we show $\frac{\Delta \alpha}{\alpha}$ over the whole sky at three different redshifts for model E. As we go back in time a larger fraction of the sky obtains $\alpha \simeq \alpha_0$. For redshifts $z > z_{\rm SSB}$ the whole sky (except very shallow voids) will have $\alpha \simeq \alpha_0$. This implies that if $\alpha$ is found to deviate from $\alpha_0$ at redshift $z_*$ then $z_{\rm SSB} > z_*$ is required for the symmetron model to be able to explain it.

\subsection*{Variation of $\alpha$ with ambient matter density}

\begin{figure}
\begin{center}
\includegraphics[scale=]{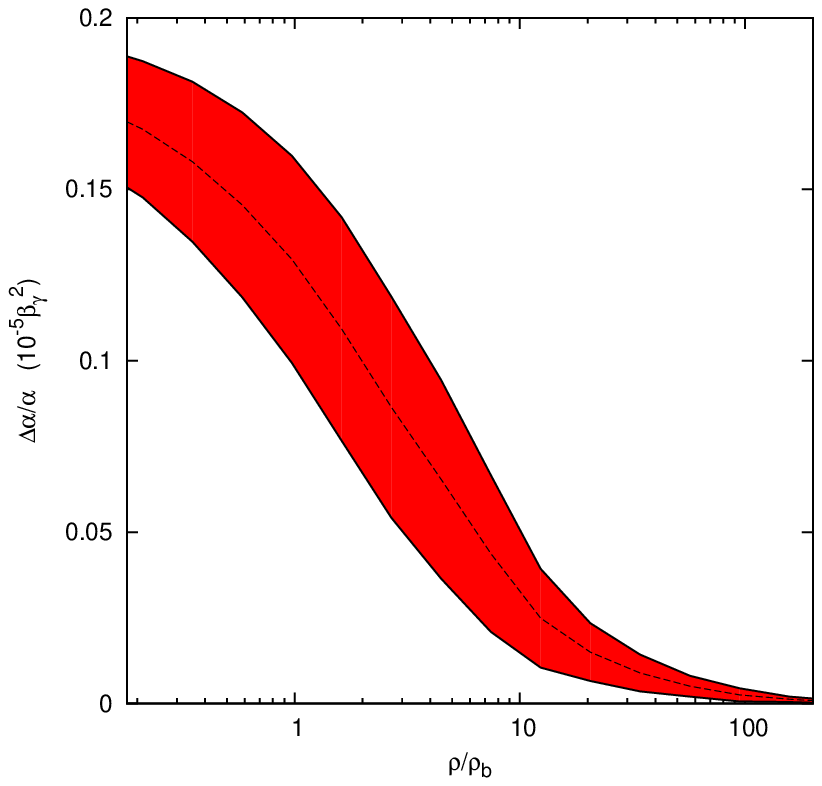}\\
\includegraphics[scale=]{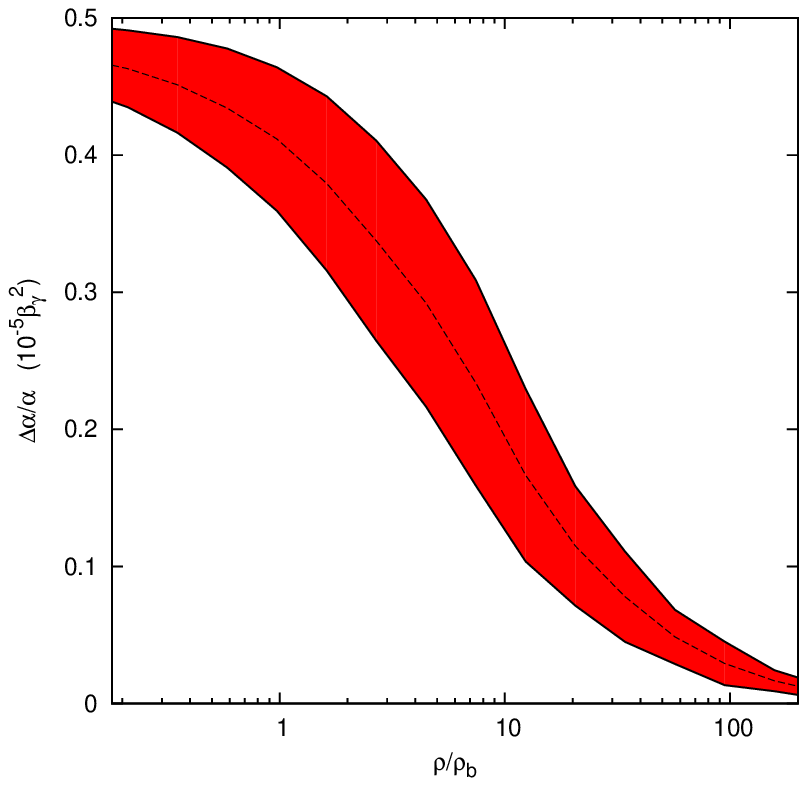}\\
\includegraphics[scale=]{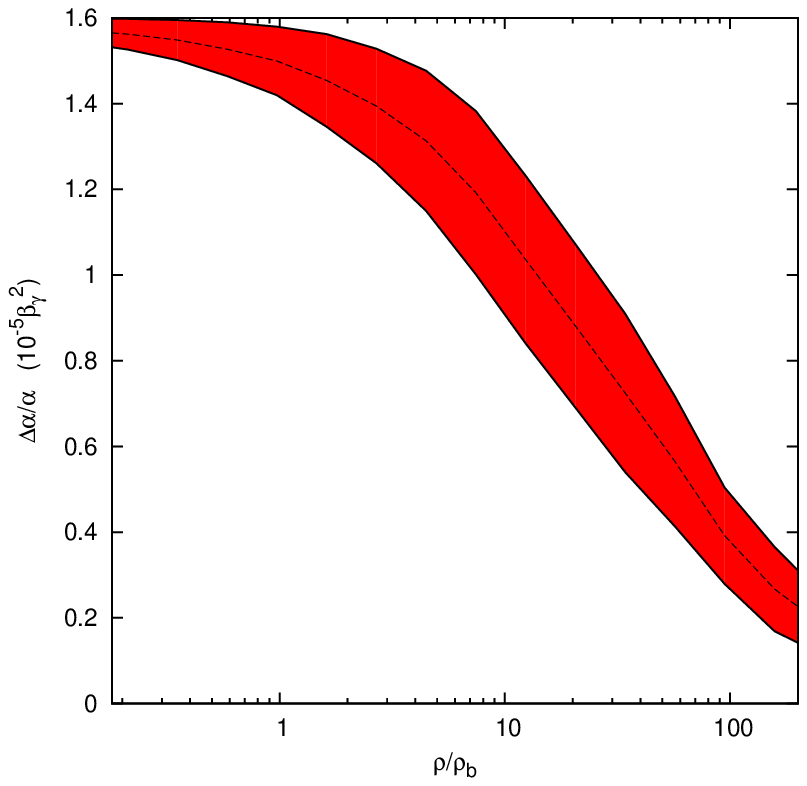}
\caption{The variation $\frac{\Delta \alpha}{\alpha}$ as function of the ambient matter density contrast for the models A (top), C (middle) and E (bottom). The spread shows the 1$\sigma$ variation from the average value. Note the different scale of the vertical axis in each case.}
\label{fig:dens}
\end{center}
\end{figure}

We investigate the correlation between the variation of $\alpha$ and the ambient matter density. For each N-body particle we calculate the ambient matter-density and the average value of $\alpha$ in a sphere of radius $r = 1$ Mpc$/h$ around the particle. The binned (in $\rho/\overline{\rho}$) result can be seen in Fig.~(\ref{fig:dens}). The spread in the figure shows the 1$\sigma$ deviation from the average in each bin.

This spread in values of $\alpha$ for any given density contrast $\rho/\overline{\rho}$ is a result of the local scalar field value depending not only on the density, but also the local environment. If we had, for example, an absorption cloud located inside a large cluster and an identical cloud at the outskirts of a cluster then in this class of models the values of $\alpha$ would differ. 

The environmental dependence of the scalar field value can have other interesting signatures. For example, in \cite{envdep} it was found that this leads to an environmental dependence on the dynamical and lensing mass estimates of dark matter halos in the symmetron model.

\section{The $\alpha$ power-spectrum}\label{alphapk}

The matter power-spectrum is a useful way to characterize the clustering scales of matter in the universe. Likewise, a  power-spectrum of $\alpha$ will track the clustering scales of the scalar-field (which determines $\alpha$). As we shall see below, the $\alpha$ power-spectrum is closely related to the matter power-spectrum for the symmetron model.

\begin{figure}
\begin{center}
\includegraphics[scale=0.9]{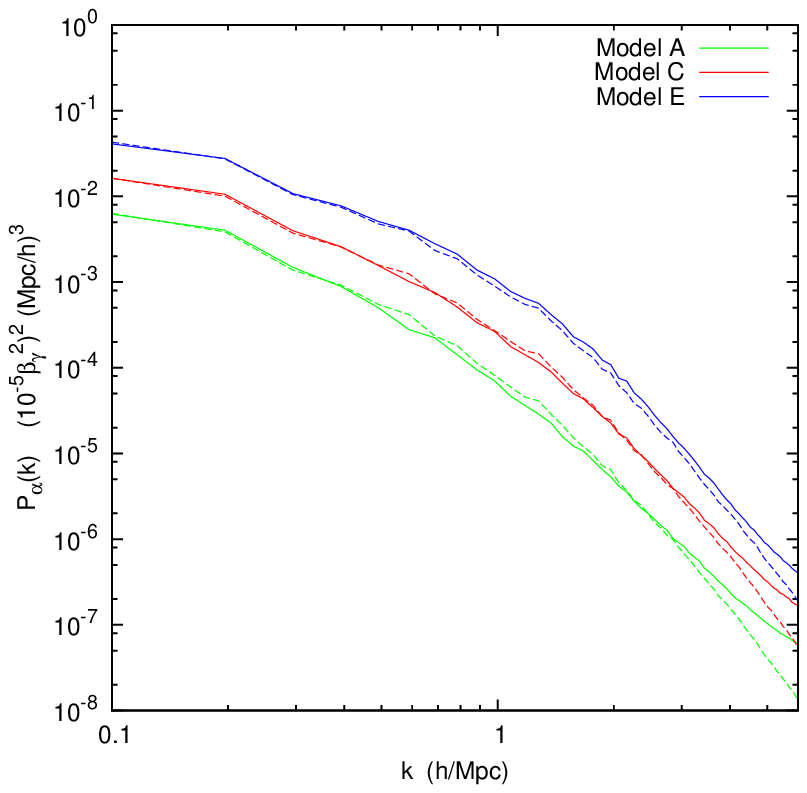}
\caption{The ($\alpha-\overline{\alpha}$) power-spectrum at $z=0$ for the models A, C and E (solid) together with the analytical expression Eq.~(\ref{apow}) (dashed).}
\label{fig:pa}
\end{center}
\end{figure}

At the linear level and in the quasi-static approximation we have that the perturbations of the scalar field in Fourier space, $\phi(k,a) = \overline{\phi}(a) + \delta\phi(k,a)$, satisfies \cite{unified}
\begin{align}
\delta\phi \simeq -\frac{\overline{\rho}_m}{M_{\rm Pl}}\frac{\beta a^2}{k^2 + a^2m_\phi^2}\left(\frac{\overline{\phi}}{\phi_0}\right)\delta_m
\end{align}
where $m_\phi^2 = V_{\rm eff,\phi\phi}(\overline{\phi})$ is the scalar field mass in the cosmological background, $\delta_m$ is the matter density contrast and $k$ is the co-moving wavenumber. The Fourier modes of $\alpha$ at linear scales then becomes
\begin{align}
\frac{\alpha(k,a)}{\alpha_0} &= 1 + \frac{1}{2}\left(\frac{\beta_\gamma(\overline{\phi} + \delta\phi)}{M}\right)^2 \simeq  \frac{\overline{\alpha}(a)}{\alpha_0} + \frac{\beta_\gamma^2\overline{\phi}\delta\phi}{M^2}\\
&=  \frac{\overline{\alpha}(a)}{\alpha_0}  - \left(\frac{\overline{\phi}}{\phi_0}\right)^2\left(\frac{\overline{\rho}_m}{M_{\rm Pl}^2}\frac{\beta_\gamma^2\beta^2a^2}{k^2 + a^2m_\phi^2}\delta_m\right)
\end{align}
where $\overline{\alpha}(a) \equiv \alpha_0\left(1+ \frac{1}{2}\left(\frac{\beta_\gamma\overline{\phi}(a)}{M}\right)^2\right)$ is the value of $\alpha$ corresponding to the scalar field value in the cosmological background. To construct a power-spectrum of $\alpha$ it is convenient to compare $\alpha(k,a)$ relative to $\overline{\alpha}(a)$ since
\begin{align}\label{aexpr}
\frac{\alpha(k,a) - \overline{\alpha}(a)}{\alpha_0} \simeq  -\beta_\gamma^2\beta^2\frac{3\Omega_m}{a}\frac{H_0^2}{k^2 + a^2m_\phi^2}\delta_m
\end{align}
is directly proportional to the matter perturbation $\delta_m$. We therefore define
\begin{align}
P_{\alpha-\overline{\alpha}}(k,a) \equiv \left|\frac{\alpha(k,a) - \overline{\alpha}}{\alpha_0}\right|^2
\end{align}
Using Eq.~(\ref{aexpr}) we find
\begin{align}\label{apow}
P_{\alpha-\overline{\alpha}}(k,a) =  \left[\frac{3\Omega_mH_0^2\beta_\gamma^2\beta^2}{a(k^2 + a^2m_\phi^2)} \left(\frac{\overline{\phi}}{\phi_0}\right)^2\right]^2P_m(k,a)
\end{align}
where $P_m(k,a) = |\delta_m(k,a)|^2$ is the matter power-spectrum. The background field value and the scalar field mass is given by \cite{symmnbody}
\begin{align}
\left(\frac{\overline{\phi(a)}}{\phi_0}\right)^2 &= \left(1 - \left(\frac{a_{\rm SSB}}{a}\right)^3\right),~~~a \leq a_{\rm SSB}\\
m_\phi^2(a) &= \frac{1}{\lambda_{\phi 0}^2}\left(1 - \left(\frac{a_{\rm SSB}}{a}\right)^3\right),~~~a \leq a_{\rm SSB} 
\end{align}
and by using $H_0 = \frac{h}{2.998\cdot 10^3 \text{Mpc}}$ we get
\begin{align}
P_{\alpha-\overline{\alpha}}(k,a) =  \left[\frac{0.33\cdot\Omega_m10^{-6}\beta_\gamma^2\beta^2}{a((k/m_\phi)^2 + a^2)}\left(\frac{\lambda_{\phi 0}}{\text{Mpc}/h}\right)
^2\right]^2P_m(k,a)
\end{align}
In Fig.~(\ref{fig:pa}) we plot the $\alpha-\overline{\alpha}$ power-spectrum at the present time, calculated from our simulations, together with the analytical result above. $P_m(k)$ is taken to be the full non-linear matter power-spectrum and we have normalized the analytical result to agree with the numerical one on large scales \footnote{The normalization constant is found to be well described by $x = 0.06\cdot \left(0.5/a_{\rm SSB}\right)^3$.} The analytical result Eq.~(\ref{apow}) is based on perturbation theory, but gives a remarkably good fit (modulo a constant factor) up to  $k\sim 3~h/$Mpc which coincides with the particle Nyquist frequency of the simulation and the grid used to calculate the power-spectrum (in other words we cannot trust the results for larger wavenumbers).

This result implies that the perturbations in the scalar field track the matter perturbations very closely even in the non-linear regime. In modified gravity models with a screening mechanism such as the symmetron this sort of effect is expected as the scalar field will sit close to the minimum of the effective potential, which is determined by the local matter density, in most regions of space.

\begin{figure}
\begin{center}
\includegraphics[scale=0.9]{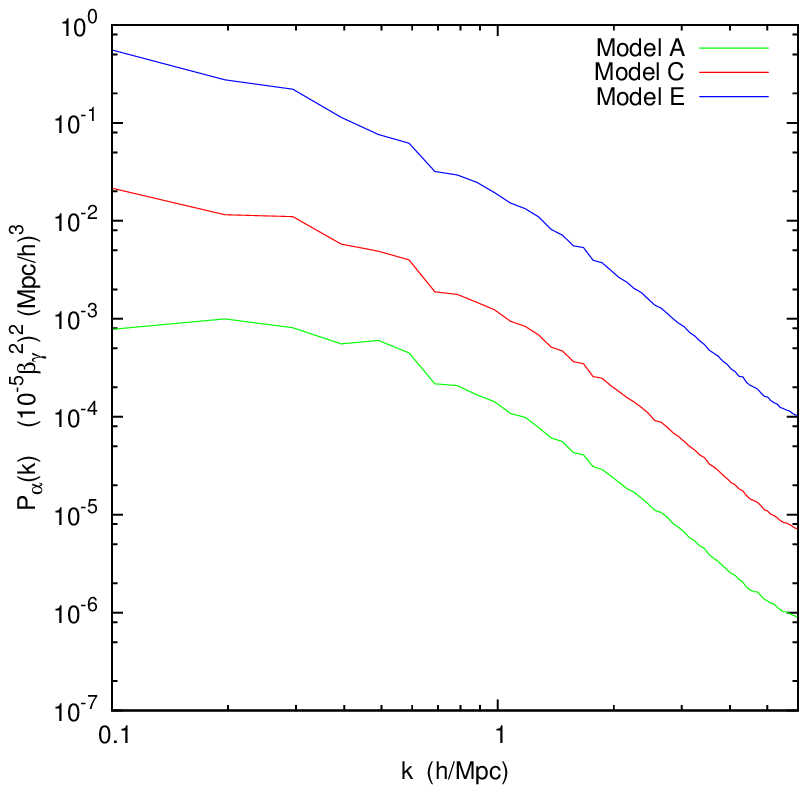}
\caption{The ($\alpha-\alpha_0$) power-spectrum at $z=0$ for the models A, C and E (solid).}
\label{fig:pe}
\end{center}
\end{figure}

For comparison, in Fig.~(\ref{fig:pe}) we show the $\alpha-\alpha_0$ power-spectrum, $P_{\alpha-\alpha_0}(k,a) \equiv \left|\frac{\alpha(k,a) - \alpha_0}{\alpha_0}\right|^2$, at the present time. As expected, an earlier symmetry breaking leads to more power.

\section{Conclusions}\label{conc}

We have investigated variations of the fine-structure constant in a particular class of scalar-tensor modified gravity model known as the symmetron. In these models the VEV of a scalar field depends on the local mass density, becoming large in regions of low density, and small in regions of high density. The coupling of the scalar to matter is proportional to the VEV and this leads to a viable theory where the scalar can couple with gravitational strength ($\beta = \mathcal{O}(1)$) in regions of low density, but is decoupled and screened in regions of high density. By coupling the scalar field to the electromagnetic field-strength tensor a spacetime variation of the scalar field will then induce a variation of $\alpha$.

The scalar field approaches $\phi \approx 0$ in high-density regions of space (such as deep inside dark matter halos) and the corresponding value of $\alpha$ approaches the value measured on Earth. In the low-density outskirts of halos the scalar field value can approach the symmetry breaking value $\phi \approx \phi_0$ and leads to value of $\alpha$ different from the one we measure on Earth. If the scalar-photon coupling strength $\beta_\gamma$ is of order unity we found that the variations of $\alpha$ inside dark matter halos are at the same level as the tentative claims by Webb {\em et al.} \cite{Dipole1,Dipole2}.

Our results also show that with low-redshift symmetry breaking these models exhibit some dependence of $\alpha$ on lookback time, as opposed to a pure spatial dipole. As the analysis of Webb {\em et al.} shows, currently available data is insufficient to distinguish between these two scenarios. It is clear that it also lacks the sensitivity to probe the characteristic environmental dependence. Nevertheless, both of these signatures can in principle be detected by sufficiently accurate spectroscopic measurements, such as those of ALMA and the ELT-HIRES.

\begin{acknowledgments}
This work was done in the context of the project PTDC/FIS/111725/2009 from FCT (Portugal). C.J.M. is also supported
by an FCT Research Professorship, contract reference IF/00064/2012, funded by FCT/MCTES (Portugal) and POPH/FSE (EC).
D.F.M. and H.A.W. thank the Research Council of Norway FRINAT grant 197251/V30. H.A.W. thanks Sigurd N{\ae}ss for help with making the skymaps. D.F.M. is also partially supported by projects CERN/FP/123618/2011 and CERN/FP/123615/2011. 
\end{acknowledgments}
\newpage
\bibliography{notes}

\end{document}